\documentclass[a4paper,11pt,headings=big,DIV=12]{scrartcl}
\usepackage{amsmath,amssymb}
\usepackage{color,xcolor}
\definecolor{blue}{rgb}{0,0,0.5} 
\usepackage[colorlinks,linkcolor=blue,urlcolor=blue,citecolor=blue]{hyperref}
\usepackage{graphicx}
\addtokomafont{disposition}{\rmfamily\boldmath}
\usepackage[utf8]{inputenc}
\usepackage{enumerate} 
\usepackage{slashed}
\usepackage{cite} 
\usepackage{multirow}
 \usepackage{bbold}
\allowdisplaybreaks
\newcommand{\W}{W}
\newcommand{\SW}{S_{\rm W}}

\newcommand{\ba}{a}
\newcommand{\bb}{b}
\newcommand{\bc}{c}

\newcommand{\bcp}{c'}
\newcommand{\baS}{a^{\rm free}_{(0)}}

\newcommand{\vev}[1]{\langle #1 \rangle}

\newcommand{\al}{\alpha}

\newcommand{\ga}{\gamma}
\newcommand{\de}{\delta}
\newcommand{\la}{\lambda}

\newcommand{\ff}{s}

\newcommand{\pd}[1]{\frac{\partial}{\partial #1}}
\newcommand{\dd}[1]{\frac{d}{d #1}}

\newcommand{\ddb}[2]{\frac{d #2}{d #1}}
\newcommand{\pdb}[2]{\frac{\partial #2}{\partial #1}}

\newcommand{\NN}{{\cal N}}

\newcommand{\Rtr}[2]{\Theta}
\newcommand{\Rtrex}[2]{#1_{\phantom{x} #2}^{#2}}

\newcommand{\til}[1]{\tilde{#1}} 
\newcommand{\dott}[1]{\partial_{\,\ln\mu}{#1}}
\newcommand{\Gt}{\W_{\tau}}
\newcommand{\GaWZ}{S_{\rm WZ}}
\newcommand{\UV}{\rm UV}
\newcommand{\IR}{\rm IR}
\newcommand{\baIR}{ \ba(\mu_{\IR} )  }
\newcommand{\baUV}{ \ba(\mu_{\UV})   }

\newcommand{\LUV}{\Lambda_{\UV}}
\newcommand{\Qmu}[1]{[{\mathcal{D}} #1]}
\newcommand{\ZZ}{\hat{ Z}}
\newcommand{\tr}{{\rm tr}}
\newcommand{\Tr}{{\rm Tr}}
\newcommand{\Dt}[1]{ D^{(#1)} }
\newcommand{\Wal}{\al}
\newcommand{\gQCD}{g }
\newcommand{\gm}{\mathrm{g}}

\begin{document}

\begin{flushright}
\begin{tabular}{l}
CP3-Origins-2015-042 DNRF90 \\
DIAS-2015-42
\end{tabular}
\end{flushright}
\vskip1.5cm

\begin{center}
{\large\bfseries \boldmath ${\cal N} = 1$ Euler Anomaly Flow from Dilaton Effective Action
}\\[0.8 cm]
{\Large%
Vladimir Prochazka
and Roman Zwicky,
\\[0.5 cm]
\small
 Higgs Centre for Theoretical Physics, School of Physics and Astronomy,\\
University of Edinburgh, Edinburgh EH9 3JZ, Scotland 
} \\[0.5 cm]
\small
E-Mail:
\texttt{\href{mailto:v.prochazka@ed.ac.uk}{v.prochazka@ed.ac.uk}},
\texttt{\href{mailto:roman.zwicky@ed.ac.uk}{roman.zwicky@ed.ac.uk}}.
\end{center}

\bigskip
\pagestyle{empty}

\begin{abstract}\noindent
We consider $\NN=1$ supersymmetric gauge theories in the conformal window. 
The running of 
the gauge coupling is absorbed into the metric  by applying a suitable matter superfield-  and 
Weyl-transformation.  The computation becomes equivalent to one of 
a free theory in a curved background carrying the information of the renormalisation group 
flow. 
We use  the techniques of conformal anomaly matching and  dilaton 
effective action, by Komargodski and Schwimmer,  to rederive the  difference of the Euler anomaly coefficient $\Delta a \equiv a_{\UV} - a_{\IR} $ 
for the $\NN=1$ theory.  
The structure of $\Delta a $ is therefore in one-to-one correspondence with the Wess-Zumino dilaton action.

\end{abstract}


\newpage

\setcounter{tocdepth}{2}
\setcounter{page}{1}
\tableofcontents
\pagestyle{plain}

\section{Introduction}
\label{sec:intro}

An exact expression for the difference of the ultraviolet (UV) and infrared (IR)  Euler anomaly
$\Delta a \equiv a_{\UV} - a_{\IR} $  was derived for  $\NN=1$ supersymmetric gauge theories
by Anselmi, Freedman, Grisaru, and Johansen (AFGJ)
 \cite{AFGJ97}.
 Thereafter it has served as  a fruitful laboratory for testing  different techniques by rederiving 
 the result. Examples include verification up to fourth loop order  \cite{JP14},  the use  
 of the local renormalisation group (RG)   \cite{BKZR14} and employing  
 superspace techniques  
 assuming  a gradient flow equation \cite{AKZ15}. In the latter case an  
  expression valid outside the fixed point has been obtained  \cite{AKZ15} of a form 
conjectured earlier by a perturbative approach \cite{FO98}.

In this paper  $\Delta  \ba |_{\NN=1}$ is derived by using  the  techniques of conformal anomaly matching and dilaton effective action. 
The latter were used by Komargodski and Schwimmer  (KS) \cite{KS11,K11} to derive 
the a-theorem $\Delta a  \geq 0$ as conjectured in 1988 by Cardy \cite{C88}.  
A crucial ingredient is the introduction of 
an \emph{external} field called the dilaton by coupling it to the renormalisation scale 
$\mu \to \mu e^{\tau(x)}$, thereby introducing a local scale interpretation analogous to  
the the local RG  pioneered by Shore \cite{Sh86,O91}.
The locality of the approach is crucial and served Jack and Osborn to derive a proof  the a-theorem 
at weak coupling (i.e. perturbation theory) by using it as a source term in a field theory in a generic curved background. KS and later Komargodski  \cite{K11} focused on the four point dilaton function and were able to prove the a-theorem based on analyticity assumptions. 
In essence the dilaton serves as a compensator field to the 
Weyl-rescaling 
\begin{equation} 
\label{eq:Weyl-g}
\gm_{\mu \nu} \to e^{-2 \Wal(x)} \gm_{\mu \nu} \;.
\end{equation}
The transformation \eqref{eq:Weyl-g} corresponds to changing  distances locally and implies that coordinate and momenta invariants change as $ x^2 \to e^{-2 \Wal(x)} x^2$ and $p^2 \to e^{2 \Wal(x)} p^2$.
Variation of the logarithm of the partion function with respect to the Weyl-parameter results in the 
vacuum expectation value (VEV) of the trace of the energy momentum tensor  (EMT). 
For a theory on a curved space,  with no explicit scale symmetry breaking, the 
EMT is parametrised by \cite{CD78,20years}
\begin{equation} 
\label{eq:TA}
\vev{ \Rtr{T}{\rho} } =  \ba \, E_4   + \bb\, W^2 + \bc \, H^2 + \bcp \Box H \;,   
\end{equation}
where  the abbreviations  
\begin{equation}
\label{eq:Theta}
\Rtr{{T}}{\rho}  \equiv  \Rtrex{{T}}{\rho} \;, \quad H \equiv \frac{R}{d-1}  \;,
\end{equation}
are used throughout. 
The quantities $E_4 = R_{\mu \nu \alpha \beta}^2 - 4 R_{\mu \nu}^2 + R^2$, 
$W^2= R_{\mu \nu \alpha \beta}^2 - \frac{4}{(d-2)}  R_{\mu \nu}^2 + \frac{2}{(d-1)(d-2)} R^2$ and $R$ are 
the Euler density, the Weyl tensor squared and the Ricci-scalar; and $R_{\mu \nu \alpha \beta}$ 
and $R_{\mu \nu}$ denote the Riemann and Ricci tensors.
The Euler density $E_4$ is a topological quantity and
the Weyl tensor squared $W^2$ vanishes on a conformally flat space.
The absence of $\bc$, and therefore the $H^2$-term, in a 4D conformal field theory (CFT) was 
shown in Ref.~\cite{BCR83}. 
The $\Box R$ term can be removed by 
a finite $H^2$-counterterm  in the action \cite{CD78} and will therefore not be discussed 
any further throughout.
The constants $\ba,\bb,\bc$ and $\bcp$ depend on the dynamics of the theory. 
Their free field values for various spins were computed in \cite{CD78}.  
Note, the non-vanishing of $\ba$ and $\bb$ therefore establish the conformal or 
Weyl anomaly in 4D \cite{CD78,20years}.

This work is structured as follows. In section \ref{sec:frame}  the general framework is outlined 
by restating some of the results of  \cite{KS11} in a language appropriate for this work. 
The specific construction  is presented and illustrated in sections 
\ref{sec:na} and  \ref{sec:toy} respectively. In section \ref{sec:N=1}, in particular \ref{sec:N=1,comp},
the AFGJ Euler anomaly result is rederived within our framework using the Konishi anomaly.
The paper ends with conclusions in section \ref{sec:conclusions}. 
Appendices  \ref{sec:Konishi}, \ref{app:free} and \ref{app:RGs}   include a review of the derivation of the NSVZ beta function using the Konishi anomaly, 
the derivation of the 
 the trace anomaly for a free theory for a  dilaton background field without  additional background  curvature  and renormalisation group equations for the generating functional.

\section{General framework}
\label{sec:frame}

Consider a massless theory with   fields $\phi$ and a 
 coupling $g$. 
 The path integral is given by 
\begin{equation}
\label{eq:partition}
 e^{ \W(g(\mu),\mu) }   =   \int \Qmu{\phi}_\mu \ e^{- \SW( g(\mu),\mu,\phi)}  \;,
\end{equation}
where the action  $\SW$ is to be understood in a Wilsonian sense and $\W$ is proportional to 
the negative free energy. 
For the purposes of this work $\SW$ is interpreted to be on a renormalisation trajectory from the UV to an IR fixed point.

In massless theories correlation functions  depend on ratios of
${q^2}/{\mu^2}$ where $q$ denotes an external momentum.
Hence the renormalisation scale transforms as 
$\mu \to e^{-\Wal} \mu$ under the Weyl-rescaling. 
An \emph{external} field, known as the dilaton $\tau$, is introduced in the action 
 \begin{equation} 
\label{eq:Seff-tau}
 \SW(g(\mu), \mu, \phi) \to    S_\tau    \equiv  \SW(g(\mu e^{ \tau}), \mu e^{ \tau}, \phi )  
   \;. \end{equation}
 transforming under Weyl-rescaling \eqref{eq:Weyl-g} as 
\begin{equation}
\label{eq:Weyl-tau}
\tau \to \tau + \Wal \;,
\end{equation}
such that the product  $\mu e^\tau$ is Weyl-invariant.  The dilaton therefore serves as a spurion (or compensator) formally restoring scale invariance.
In this work no dynamic nature is attributed to the dilaton field which is in line with \cite{K11} but not 
the first paper \cite{KS11} on the a-theorem in 2011. 
 The dilaton  serves as a  source term for the EMT and when made a local field the, yet to-be-defined,  
 Wess-Zumino term carries the information on the Euler anomaly. 
 Promoting the dilaton to  a local field  $\tau \to \tau(x)$  requires  
  \emph{local}  Weyl invariance and demands  changes similar to passing from global to  local gauge invariance. The specific implementation will be discussed in the the explicit examples.
 The space-dependance of $\tau$  augments the couplings to local objects  $g(\mu) \to  g(\mu e^{\tau(x)})$. 
 Note that the functional form $\mu e^{\tau}$ 
 renders local Weyl-rescaling equivalent to a local RG transformation.
The  path integral becomes $\tau$-dependent,
\begin{equation} 
\label{eq:taupath}
 e^{   \W_\tau }= \int \Qmu{\phi}_\mu  e^{- S_\tau } =  \int \Qmu{\phi}_\mu
 e^{- \SW(g(\mu e^{ \tau}), \mu e^{ \tau}, \phi ) }  \;.
\end{equation}
The quantity  $\W_\tau$ corresponds to  the generating functional  of the correlation function (connected component)  
of the traces of the EMT.
The Wess-Zumino action\footnote{The    Wess-Zumino action, above, is analogous to the Wess-Zumino  term \cite{WZ71}  of pions in connection with the the axial anomaly.}
\begin{equation}
\label{eq:gaWZ}
\GaWZ =   \int d^4 x \, 2 \left(  2\Box \tau (\partial \tau)^2- (\partial \tau)^4 \right) + {\cal O}(R)  \;,
\end{equation}
was shown to be the source term   \cite{KS11} of the Euler anomaly.
Above $R$ stands for the non-dilaton curvature background.
 More precisely, using arguments of conformal anomaly matching it was shown that
the  difference of the UV and IR dilaton effective action, with with  $g^*_{(\UV,\IR)} \equiv g(\infty,0)$,
\begin{equation}
\label{eq:2}
\Delta \Gt \equiv  \int_{g^*_{\IR} }^{g^*_{\UV} } d g  \, \partial_g {\W}_\tau   =
-   \int_{-\infty}^{\infty} d \ln \mu \, \dott{\W}_\tau =
\Gt( g^*_{\UV}) - \Gt(g^*_{\IR}) = - \Delta \ba \ \GaWZ + \dots \;,
\end{equation}
contains a term  proportional to   $\GaWZ$ times the sought after quantity 
$ \Delta \ba \equiv  \baUV - \baIR $ \cite{KS11}.
Hence determining $\Delta \ba$ reduces to finding $\dott{\W}_\tau $. 
Note that the second equality in \eqref{eq:2} follows from \eqref{eq:RGE} in the limit $m \to 0$ 
and using $dg/\beta = d \ln \mu$.

\subsection{Dilaton dependent conformal factor}
\label{sec:na}

In this work a theory is considered which can be reinterpreted as a free field theory 
in a conformally flat  background 
\begin{equation}
\label{eq:gt}
\til{\gm}_{\rho \la} =  e^{-2\ff(\tau)} \de_{\rho \la} \;,
\end{equation}
which carries the information on the RG flow parameters.
Above $\delta_{\rho \la}$ denotes the flat Euclidian metric with positive signature. 
The adaption to Minkowski space is straightforward as it results in the appearance 
of various factors of $i$ only.  
By using \eqref{eq:use} from the appendix\footnote{Note that  metric \eqref{eq:gt} 
is not a physical or geometric metric as it does not transform 
like \eqref{eq:Weyl-g} under Weyl-rescaling    $s(\tau) \to s(\tau+ \al)$  unless  $\ff(x) = x$.  The latter is the case in  \cite{KS11},   $\hat \gm_{\rho \la} = e^{-2 \tau(x)} \de_{\rho \la}$, and constitutes one of the differences with respect to our approach.}\,\footnote{The subscript $\tau$ refers to the  VEV of the trace of the EMT with respect to the partition function \eqref{eq:taupath}.} 
\begin{equation}
\label{eq:1}
\dott{\W}_\tau  =   \int d^4 x  \sqrt{ \tilde \gm} \ \vev{ \Rtr{{T}}{\rho}}_{\tau} \;,
\end{equation} 
and \eqref{eq:2} we may write a  more explicit formula for the difference of the Euler  anomaly 
$\Delta a$
\begin{equation}
\label{eq:4}
\Delta \ba =   \int_{-\infty}^{\infty} d\ln \mu \ 
\int d^4 x \sqrt{ \tilde \gm} \vev{ \Rtr{{T}}{\rho}}_{\tau}|_{\GaWZ} \,,
\end{equation}
where where $|_{\GaWZ}$ denotes the following projection 
\begin{equation} 
\label{eq:TA-ac}
  \vev{ \Rtr{{T}}{\rho}  }_\tau  = \til{\ba} \til{E}_{4} + \til{\bc} \til{H}^2   = 
  \vev{ \Rtr{{T}}{\rho}}_{\tau}|_{\GaWZ}  \, \GaWZ  + ... \;.
  \end{equation}
The coefficients $\til{a}$ and $\til{c}$ depend on the dynamics of the theory.
The Wess-Zumino action $\GaWZ$ is defined in \eqref{eq:gaWZ} and  we follow the rule that the tilde 
denotes  geometric quantities, e.g.   $\til{E}_{4}$ and $\til{H}^2 $, 
evaluated in the   background 
metric $\til{\gm}_{\mu \nu} $.\footnote{Conformal flatness of $\til{\gm}$ implies that 
$\til{W}^2 =0$. } 
The quantity $\Delta a$ is determined once $\til{a}$ and $\til{c}$ are known.
In the next section we will discuss a very simple toy model that illustrates these ideas and will serve 
as a stepping stone for the $\NN=1$-computation.

\subsection{Weyl anomaly of free scalar field in conformally flat background}
\label{sec:toy}

We consider a scalar field theory on a  flat space,   focusing solely  
on the kinetic term
\begin{equation}
\label{eq:S0}
\SW(\mu) =  \int d^4 x Z(\mu)  \de^{\rho \la}  \partial_{\rho} \phi  
\partial_{\la} \phi \;,
\end{equation}
thereby ignoring other contributions.
The usefulness of which will, hopefully, become clear in section 
\ref{sec:N=1,comp}.
Taking $Z(\mu) \to Z(\mu e^{\tau(x)})$ amounts to passing to $S_\tau$  
\eqref{eq:Seff-tau}.
The  factor $Z(\mu e^{\tau(x)})$ can be absorbed into the metric by a  
local  Weyl-rescaling
by choosing $\Wal = s$ in \eqref{eq:Weyl-g} with
\begin{equation}
\label{eq:g-t}
  \ff(\mu e^\tau) = -\frac{1}{2} \ln Z(\mu e^{\tau(x)}) \quad  
\Rightarrow  \quad
   \til{\gm}_{\rho \la}  =  Z  \de_{\rho \la} \;.
\end{equation}
The theory then becomes a field theory on  a conformally flat space  
with metric $\til{\gm}_{\mu \nu}$ \eqref{eq:g-t}
\begin{equation} \label{eq:S0scalar}
S_\tau(\mu) =  \int d^4 x \sqrt{\til{\gm}}  \til{\gm}^{\rho \la }  
\Dt{\ff}_{\rho} \phi \Dt{\ff}_{\la} \phi \;.
\end{equation}
Above $\Dt{\ff}_\rho = \partial_\rho -(\partial_\rho \ff)$
denotes the  Weyl-covariant derivative\footnote{Adding the  
Weyl-covariant derivatives is equivalent
to the replacement $\Box \to \Box - \frac{1}{6} \til{R}$ which is the usual
conformally coupled scalar in a curved space of metric  
$\til{\gm}_{\rho \la}$.}
analogous to the covariant derivative in gauge theories.
The coefficients of the trace anomaly \eqref{eq:TA} for a theory with  
metric $\til{\gm}_{\rho \la}$
and one free scalar field are given by  (cf.  
\cite{CD78,birrell1982quantum} or the explicit computation in appendix  
\ref{app:free})
\begin{equation}
\label{eq:Afree0}
\til{\ba} = \baS  \;, \quad \til{\bc} =0 \;, \quad \til{c}' = - 2 \baS
  \;, \qquad \baS =
  \frac{1}{ 360} \frac{1}{16 \pi^2} = \frac{1}{5760 \pi^2} \;,
\end{equation}
or equivalently $
  \vev{ \Rtr{{T}}{\rho}  }_\tau =
\baS  ( \til{E}_4 -2 \til{\Box} \til{R}) $.
As stated earlier,
the coefficient $c'$ is of no importance for this work and is  
therefore discarded.
The Euler density
in terms of $\ff$ is given by
\begin{equation}
\label{eq:E4}
\sqrt{\til{\gm}} \til{E}_4=
- 8 ( \frac{1}{2}\Box (\partial \ff)^2 - \partial \cdot(\partial  
\ff(\Box \ff - (\partial \ff)^2))) \;.
\end{equation}
  Using the explicit form $\ff = -\frac{1}{2} \ln Z(\mu e^\tau)$ the  
Euler term becomes
\begin{eqnarray}
\label{eq:E4-ga2}
\sqrt{\til{\gm}} \til{E}_4= -[  \ga^2  \Box (\partial \tau)^2 +(2\ga  
\dot{\ga} - 2\ga^2)  \partial^{\la}(\partial_{\la} \tau \Box  
\tau) - \ga^3  \partial^{\la}(\partial_{\la} \tau (\partial  
\tau)^2)  - \nonumber  \\[0.1cm] 6\ga \dot{\ga} (\partial \tau)^2 \Box \tau
-3 \ga^2 \dot{\ga} (\partial \tau)^4 \big ] \;,
\end{eqnarray}
where here and below we use the abbreviation  $\dot{\ga} \equiv  
\frac{d}{d \log \mu}{\ga}$ and the following expressions
\begin{equation} \label{TildeDerivative}
\partial_\rho \ga =  \dot{\ga} \, \partial_{\rho}\tau   \;,  \quad
  \partial_{\rho} \ff =  - \frac{1}{2} \frac{\partial \ln Z(\mu  
e^{\tau}) }{\partial (\mu e^{\tau})}
  \partial_\rho (\mu e^{\tau} )  =  - \frac{1}{2} \ga \,  
\partial_{\rho}\tau \;, \quad
\ga   =  \frac{\partial \ln Z(\mu) }{\partial \ln \mu}  \;,
\end{equation}
have been used.
The quantity $\Delta \ba$  is obtained by integrating over $d \ln \mu$  
and projecting on
  $\GaWZ$. In doing so
  $\ga$ and $\dot{\ga}$ can be treated as being space-independent,
since expanding  $\gamma(\mu e^{\tau})= \gamma(\mu)+ O(\tau(x))$ leads  
to terms which are not contained in
$\GaWZ$.
Furthermore  it is then clear that the first line in \eqref{eq:E4-ga2}  
can be discarded since it is a total
derivative and therefore inequivalent  to the  $\GaWZ$ \eqref{eq:gaWZ}  
bulk-term.  In order to project the second line of \eqref{eq:E4-ga2}  
on $\GaWZ$ \eqref{eq:gaWZ}
it is convenient (following \cite{KS11},\cite{K11})\footnote{We note in passing that Eq.~\eqref{eq:eom-t} is the lowest order equation of motion 
\cite{KS11} when a dynamic nature is attributed the dilaton.} \begin{equation}
\label{eq:eom-t}
\Box \tau = (\partial \tau)^2 \;,
\end{equation}
under which all four-derivative invariants vanish, except for
\begin{equation}
\label{eq:GaWZeom}
\GaWZ|_{ \eqref{eq:eom-t}}  = \int d^4 x  \, 2 (\Box \tau)^2 \,.
\end{equation}
Using \eqref{eq:4} and performing the integral over $d \ln \mu$ we get
\begin{equation}
\label{eq:4DilatonTerm}
  \Delta a =    \frac{1}{2} \baS
   \big [3A_1 + A_2 ] \;,
\end{equation}
where
\begin{alignat}{3}
\label{DeltaGammaCoeff} \notag
A_1 &= \int _{- \infty} ^{\infty} d \ln \mu \ 2\ga \dot{\ga} &\;=&  
\int _{\ga_{\IR}}^{\ga_{\UV}} d \ga \ 2 \ga &=\;& (\ga_{\UV}^2 -  
\ga_{\IR}^2)  \;,\\
A_2 &= \int _{- \infty} ^{\infty} d \ln \mu \ 3 \ga^2  
\dot{\ga}&\;=&\int _{\ga_{\IR}}^{\ga_{\UV}} d \ga \ 3 \ga^2 &=\;&  
(\ga_{\UV}^3 - \ga_{\IR}^3) \;,
\end{alignat}
and $\ga_{\IR,\UV} \equiv \ga( g^*_{\IR,\UV}) $ are the values of the anomalous dimensions at the respective fixed points.
 For further reference the final result \eqref{eq:4DilatonTerm} is stated with explicit coefficients 
$A_1$ and $A_2$
\begin{eqnarray}
\label{eq:Da-free}
\Delta \ba    
&=&   \frac{1}{2}\left(
(\ga_{\UV}^3 - \ga_{\IR}^3)  + 3 (\ga_{\UV}^2 - \ga_{\IR}^2) \right) \baS \;.
\end{eqnarray}
This result constitutes an important intermediate result for the derivation of $\Delta a|_{ \NN=1}$.

\section{$\NN = 1$ supersymmetric gauge theory}
\label{sec:N=1}

The theory considered in this section is  a $\NN=1$ supersymmetric gauge theory with flavour symmetry $SU(N_f) \times SU(N_f)$ and gauge group $SU(N_c)$. 
The  action can be written in terms of the usual vector superfield $V$ and matter superfields $(\Phi_{f},\tilde{\Phi}_f)$ as, e.g. \cite{S-book}, 
\begin{equation}
\label{eq:S,N=1}
 \SW(\mu) = \int d^6 z \frac{1}{\gQCD^2(\mu)} \tr W^2 + {\rm h.c.} + 
 \frac{1}{8} Z(\mu) \sum_{f} \big [\int d^8 z    \Phi_{f}^\dagger e^{-2 V} \Phi_{f} + \int d^8 z    \tilde{\Phi}_{f}^\dagger e^{-2 V}  \tilde{\Phi}_{f} \big ]\;,
\end{equation}
where $W^2$ is the supersymmetric gauge field kinetic term,  $\gQCD$ is referred to  as the holomorphic  coupling constant parametrisation and 
$d^6 z $ and $d^8 z$ include integration over the fermionic superspace variables. 

The main tool in deriving $\Delta  \ba |_{\NN=1}$ is the use of the Konishi anomaly \cite{K83,KS85,ShV86}. The latter is illustrated in appendix \ref{sec:Konishi} 
as a method to derive the NSVZ beta function. 
In section  \ref{sec:N=1,comp} the Konishi anomaly is used 
 to write the Wilsonian action such that the RG flow   can be absorbed into the metric.  
This procedure  makes  it  amenable to the free field theory computation  
in the dilaton background discussed in section \ref{sec:toy}.

\subsection{$\Delta  \ba |_{\NN=1}$ from Dilaton effective Action and Konishi anomaly }
\label{sec:N=1,comp}

We consider the $\NN=1$ supersymmetric gauge theory with Wilsonian effective action given 
in \eqref{eq:Seff,N=1}.  Choosing a rescaling factor, with   $ \ga_* = -b_0/N_f$ \eqref{eq:gstar}, 
\begin{equation}
\label{eq:rescaling2}
(\Phi_{f},\tilde{\Phi}_f) \to  \left( \frac{\mu'}{\mu}\right) ^{\ga_*/2} (\Phi_{f},\tilde{\Phi}_f)
\end{equation}
on the matter fields the Konishi turns the action into the 
following form
\begin{eqnarray}
\label{eq:Sbare}
 \SW(\mu) &\;=\;& \int d^6 z \frac{1}{\gQCD(\mu')^2}  \tr W^2 + {\rm h.c.} +  \nonumber \\[0.1cm]
&\;\phantom{=}\;& \frac{1}{8} \sum_{f} \big [\int d^8 z \hat{Z}(\mu)   \Phi_{f}^\dagger e^{-2 V} \Phi_{f} + \int d^8 z  \hat{Z}(\mu)  \tilde{\Phi}_{f}^\dagger e^{-2 V}  \tilde{\Phi}_{f} \big ]\;,
\end{eqnarray}
with
\begin{equation}
\ZZ (\mu) \equiv Z(\mu) \left( \frac{\mu'}{\mu} \right)^{\ga_*} \;,
\end{equation}
where $\mu' > \mu$ is an arbitrary scale which can be thought off as a UV cut-off $\LUV$.
Crucially, the RG flow is absorbed into the precoefficient  $\ZZ (\mu) $ in front of the matter term. 
Eq.~\eqref{eq:Sbare} is the analogue of the action \eqref{eq:S0} for the scalar field to the degree
that the running of the theory is parametrised by a coefficient in front of the matter kinetic term.

Again following the procedure in \eqref{eq:Seff-tau}  a dilaton is introduced through 
\begin{equation}
\ZZ(\mu)  \to \ZZ_\tau(\mu)  = \ZZ(\mu e^{\tau(x)}) =
Z(\mu e^{\tau(x)}) \left( \frac{\mu'}{\mu e^{\tau(x)}} \right)^{\ga_*}  \;.
\end{equation}
To keep the procedure manifestly supersymmetric, following \cite{ST10}  the dilaton is promoted to a (chiral) superfield $T$ such that
\begin{equation}
\label{LowestZ}
 T | = \tau+ i \omega   \;, \quad \ZZ(\mu e^T) | = \ZZ(\mu e^{\tau}) \;.
\end{equation}
Above $\omega$ is the axion and the bar stands for projection on to the lowest component of the multiplet.
It will be seen that $\ZZ$  in \eqref{eq:Sbare} 
 can be absorbed into the background geometry by a local Weyl-rescaling. 
 To preserve local SUSY invariance the Weyl transformations are  
promoted to super-Weyl transformations.  
Under  the latter, the $\tr W^2$-term is invariant whereas
 the matter term transforms as follows (c.f \cite{Wess:1992cp})
 \begin{equation}
 \label{eq:weyl-matter}
  \int d^8 z   \Phi^\dagger e^{-2 V} \Phi \to   \int d^8 z e^{- A } e^{ -A^{\dagger}}   \Phi^\dagger e^{-2 V} \Phi \;,
 \end{equation}
 with the superfield $A=\alpha+i \beta + \dotsc$ being the super-Weyl parameter corresponding to $\al$ in \eqref{eq:Weyl-g}.
Note that such a formalism is automatically local Weyl-invariant and that there is no need to introduce the Weyl-covariant derivatives as in \eqref{eq:g-t}. 
Furthermore, the transformation \eqref{eq:weyl-matter} amounts to a 
Weyl-rescaling of the vielbein\footnote{Under a super-Weyl transformation, the supersymmetric generalization of vielbein transforms as
$E_{\mu}^{a} \to e^{- \frac{A}{2} } e^{ -\frac{A^{\dagger}}{2}} E_{\mu}^{a}$, which corresponds to the standard Weyl transformation $e_{\mu}^{a} \to e^{-\al} e_{\mu}^{a}$ after projecting on the 
lowest component of $E_{\mu}^{a}$. In the interest of clarity we would like to add that  $e^a_\mu = \de^a_\mu$ on flat 
space.}
\begin{equation}
 \label{eq:SUWeyl}
e^a_\rho \to e^{- \frac{A}{2} } e^{ -\frac{A^{\dagger}}{2}}| \ e^a_\rho= e^{-\alpha} e^a_\rho \;.
\end{equation}
Upon identifying $\al = s$ in Eq.~\eqref{eq:g-t}  
\begin{equation}
 \label{eq:WeylSUSY}
e^a_\rho \to \til{e}^a_\rho = \sqrt{ \ZZ_\tau(\mu)}e^a_\rho \;.
\end{equation}
The action $S_W(\mu e^{T})$ \eqref{eq:Sbare} can then be written in a manifestly locally supersymmetric form; cf. section 6.3 in \cite{Buch98}.
 Eq.~\eqref{eq:WeylSUSY}  results in
\begin{equation}
\label{eq:Ztil}
\til{\gm}_{\rho \la} = \til{e}^a_\rho  \til{e}^a_\la = e^{-2\ff(\mu e^\tau)}  \de_{\rho \la} \;, 
\quad \ff(\mu e^\tau) = -\frac{1}{2} \ln \ZZ(\mu e^{\tau(x)}) \;.
\end{equation}
Notice that the UV scale $\mu'$ is arbitrary and that therefore  
a physical quantity like  $ \vev{ \Rtr{{T}}{\rho}  }_\tau$  should not depend on it.
Since the geometric terms   $\til{E}_{4}$ and 
$\til{H}^2$ are independent of $\mu'$,\footnote{To see this  notice that these terms  depend on derivatives of $s$ only (c.f \eqref{eq:E4}). The latter are related to the anomalous dimension $\gamma(\mu e^{\tau})$ through the relation \eqref{TildeDerivative} 
which is independent of $\mu'$.}  the form of \eqref{eq:TA-ac} 
implies that $\til{a}$ and $\til{c}$  are $\mu'$-independent and  therefore constants.  
This means that $\til{a}$ and $\til{c}$ assumed the values at the (free)  UV fixed point 
 and the geometric quantities are to be evaluated in the background metric $\til{g}_{\rho \la}$ carrying the dynamic information. This allows to recycle, in large parts,  the computation in   section  \ref{sec:toy} as outlined below.

A free theory in a curved background is  in particular  conformal and 
 therefore free of the $\til{R}^2$-term (i.e. $\til{c} =0$).
Since the dilaton couples to the matter part only, 
the trace anomaly is  exhausted by the free field theory computation of the matter-fields 
in the curved background with metric \eqref{eq:Ztil}. 
Equivalence to the  example in the previous section is achieved 
through the formal  replacement  $Z \to \hat{Z}$ (following from \eqref{eq:Ztil}) which implies 
$\ga \to \de \ga \equiv \ga - \ga_*\,$ \emph{and} 
the change in the number of degrees of freedom $\nu$. 
More precisely the matter superfield consists of a complex scalar and a Weyl fermion which contribute 
 \cite{CD78}
\begin{equation}
\nu \equiv 2\Big|_{\mathbb{C}\text{-scalar}} + \frac{11}{2}\Big|_{\text{Weyl-fermion}}   = \frac{15}{2}
\end{equation}
 in units of a real scalar field. This number has to be   multiplied by 
the number of colours  $2 N_f$ (two matter-field per flavour) and $N_c $ (the $SU(N_c)$ Casimir of 
the adjoint representation).  
Hence $\Delta a$ is given 
by $ 2 N_f N_c\nu \Delta a|^{\eqref{eq:Da-free}}_{\ga_{\UV,\IR}  \to  \de \ga_{\UV,\IR}}$. 
Now, $(\ga_{\UV},\ga_{\IR}) =  (0,\ga_*)$ implies $(\de \ga_{\UV},\de \ga_{\IR})  = (- \ga_*,0)$   and therefore 
\begin{equation}
\label{eq:N=1}
\Delta  \ba |_{\NN=1} =  \frac{15}{2}  N_c N_f (- \ga_*^3  + 3 \ga_*^2) \baS \;.
\end{equation}
We note that \eqref{eq:N=1}  is indeed the same as the non-perturbative result 
quoted in (Eq.4.18) in \cite{AFGJ97} when taking into account the explicit form 
of  $\ga_*$ \eqref{eq:gstar}. The formula above is valid in the conformal window 
$3/2 N_c < N_f < 3 N_c$  where the UV theory is asymptotically free 
and the IR theory acquires a non-trivial fixed point.  Within these boundaries the anomalous dimension 
$\ga_*$ takes on the values $-1$ to $0$ and the quantity $\Delta a $ is therefore manifestly positive 
in accordance with the a-theorem.  The latter has been proven for  $\NN=1$ supersymmetric theories 
by using $R$-symmetries and is known as $a$-maximization \cite{IW03}.
The adaption  to
gauge groups other than 
$SU(N_c)$, provided they are asymptotically free,  amounts to replacing the $SU(N_c)$-Casimir $N_c$ by 
the corresponding Casimir of the group.

\section{Discussion and Conclusion}
\label{sec:conclusions}

In this short paper we rederived the difference of the Euler term in $\NN=1$ supersymmetric gauge theories \eqref{eq:N=1} valid in the conformal window.  
By an appropriate rescaling of  the matter superfield \emph{and} 
choosing the  Weyl-parameter $\al$ \eqref{eq:Weyl-g}  to equal  the logarithm of the  matter prefactor  \eqref{eq:Ztil}, the computation was shown to be equivalent to one of the (free) UV theory in a curved background carrying the information on the flow. 
This allowed for $\Delta  \ba |_{\NN=1}$ to be computed from the free field theory example,  in section 
\ref{sec:toy}, with 
a simple formal  replacement for $\ga_{\IR}$ and $\ga_{\UV}$. It is noted that the structure of  $\Delta \ba|_{\NN=1}$ is completely given by the Wess-Zumino term of the dilaton effective action. The aspect of matching the computation with a free theory bears some resemblance with the original AFGJ-derivation 
\cite{AFGJ97} in that independence on an RG-scale is exploited in evaluating certain quantities in the 
UV where they correspond to free field theory computations.
 An extension to the non-supersymmetric case is not straightforward because it relies on the  one-loop exactness of the rescaling anomaly in supersymmetric gauge theories.  From 
  sections \ref{sec:Konishi} and \ref{sec:N=1,comp} it is seen that an exact expression of $\Delta a$ in 
  non-supersymmetric theories is related to finding an exact beta function.  
 The Konishi anomaly is a rescaling anomaly which 
in $\NN=1$ supersymmetric theories  is, by holomorphicity, bound to the axial anomaly.
The latter is generally one-loop exact by topological protection of the axial charge. In non-supersymmetric theories there is no holomorphicity and the Konishi anomaly is an unknown function which could be determined order by order in perturbation theory.
An extension of some of the ideas in this paper  to non-supersymmetric theories, in direction other than the Konishi anomaly,
is presented elsewhere \cite{PZpro}.
  
We end the paper with remarks of the  speculative and qualitative kind. 
Reformulating a gauge theory as a free theory in a curved background is reminiscent  of
the   anti-de Sitter space/conformal field theory duality which has given rise to a lot of work and inspiration over the past two decades. 
The extension to theories with more than one relevant coupling is not immediate. 
 One might wonder whether bi-gravity, whose renormalisation group flow has been studied in 
 \cite{D13}, might be a possible avenue for a theory with two relevant couplings. A practical requirement is that the UV theory is to asymptotically free in order to retain computability.

\subsection*{Acknowledgements}

We are grateful to Roberto Auzzi, Luigi Del Debbio, Mike Duff, Franesco Sannino, 
Ian Jack and Graham Shore  for useful discussions.
VP acknowledges the support of an STFC studentship (grant reference ST/K501980/1). 
RZ acknowledges the kind hospitality of the Marseille particle physics (CNRS) where part of this work was done.

\appendix

\numberwithin{equation}{section}

\section{$\NN=1$ effective action and the Konishi anomaly}
\label{sec:Konishi}

Using arguments of holomorphicity it can be argued that the running of the coupling $\gQCD$, 
of the Wilsonian effective action of the supersymmetric gauge theory  \eqref{eq:S,N=1}, 
is one-loop exact  \cite{AM97,S-book} and reads
\begin{eqnarray}
\label{eq:Seff,N=1}
 \SW(\mu) &\;=\;& \left( \frac{1}{\gQCD^2(\mu')}  -  \frac{b_0}{8 \pi^2} \ln \frac{\mu'}{\mu}  \right)  \int d^6 z  \tr W^2 + {\rm h.c.} \nonumber   \\[0.2cm]
&\;+\;&  \frac{1}{8} Z(\mu) \sum_{f} \big [\int d^8 z    \Phi_{f}^\dagger e^{-2 V} \Phi_{f} + \int d^8 z    \tilde{\Phi}_{f}^\dagger e^{-2 V}  \tilde{\Phi}_{f} \big ]\;,
\end{eqnarray}
where $ b_0 \equiv  3N_c- N_f$ and $\mu' > \mu$ is an arbitrary scale which can be identified 
with the UV cut-off $\LUV$.
Rescaling the  matter fields  by
\begin{equation}
\label{eq:rescaling}
(\Phi_{f},\tilde{\Phi}_f) \to Z^{-1/2} (\Phi_{f},\tilde{\Phi}_f) \;,
\end{equation}
 is accompanied by 
the Konishi anomaly \cite{K83,KS85,ShV86}, and leads to the effective action \cite{AM97} 
\begin{equation}
\label{eq:NSVZ}
 \SW(\mu) = \int d^6 z \frac{1}{\gQCD^2(\mu)} \tr W^2 + {\rm h.c.} + 
 \sum_{f} \big [\int d^8 z    \Phi_{f}^\dagger e^{-2 V} \Phi_{f} + \int d^8 z    \tilde{\Phi}_{f}^\dagger e^{-2 V}  \tilde{\Phi}_{f} \big ]\;,
\end{equation}
where the running has been removed from the matter term and all the running 
is absorbed in front of the gauge field term which in this case defines the running 
 gauge coupling to be
\begin{equation}
\frac{1}{\gQCD^{2}(\mu)} =  \frac{1}{\gQCD^{2}(\mu')}  -  
\frac{1}{8 \pi^2}\left( b_0  \ln \frac{\mu'}{\mu} - N_f \ln Z \right) \;.
\end{equation} 
The $\mu'$-independence of $g(\mu)$ 
implies an RGE which solves to the holomorphic NSVZ  beta function  \cite{NSVZ83,SVZ85,SV86}
\begin{equation}
\label{eq:NSVZ}
\beta \equiv  \dd{\ln \mu} \gQCD =  - \gQCD^3  \frac{N_f}{16 \pi^2} (\ga-\ga_*)  \;.
\end{equation}
Above we have used the following notation
\begin{equation}
\label{eq:gstar}
\ga \equiv \pdb{\ln \mu}{\ln Z(\mu)}  \;, \quad    \ga_* \equiv -b_0/N_f =  1-3 N_c/N_f \;.
\end{equation}
In the range $3/2 N_c < N_f < 3 N_c$ the theory is in the so-called conformal window. 
Theories of the latter kind are
asymptotically free in the UV and flow to a non-trivial ($\ga_{\IR} = \ga_* \neq 0$) IR fixed point.  The lower bound $3/2 N_c$ follows from the unitarity bound on the scaling dimension of the 
composite squark field 
$\Delta_{\til{\overline{q}}\til{q} } = 2 + \ga_* \geq 1$ and the 
upper bound of $3N_c$ is derives form the requirement of  asymptotic freedom.

\section{Free theory trace anomaly in dilaton background field}
\label{app:free}

In this appendix the trace anomaly of a free scalar field theory \eqref{eq:S0scalar} 
is evaluated on a conformally flat background $\til{g}_{\rho \la} = e^{-2 s(x)} \de_{\rho \la}$. The path integral is Gaussian and evaluates to
\begin{equation} 
e^{\W}= \int \mathcal{D}\phi e^{-S^{(0)}}= \sqrt{\det{\Delta^{(0)}}}= \exp{\frac{1}{2} \Tr\ln{ \Delta^{(0)}}} \;,
\end{equation}
where $ \Delta^{(0)}= e^{-2 s}(-\Box+(\Box s- (\partial s)^2)$ 
is the kinetic operator obtained from \eqref{eq:S0scalar} by integration by parts.
The contribution can be evaluated using  Schwinger's formula 
\begin{equation} \label{Schwinger}
\W= \frac{1}{2} \Tr\ln{ \Delta^{(0)}}=  \frac{1}{2} \int_{0}^{\infty} \frac{dt}{t} \Tr(e^{-t  \Delta^{(0)}}) 
\;.
\end{equation}
This expression requires regularisation since it is UV-divergent  as $t \to 0$.
Noting that the mass dimension of the $t$-variable is two, a   UV cutoff $\LUV$ is introduced 
as follows
\begin{equation} \label{FreeCounter}
\W_{\rm reg}=  \frac{1}{2}  \int_{\LUV^{-2}}^{\infty} \frac{dt}{t} \Tr(e^{-t  \Delta^{(0)}}) \;.
\end{equation}
Using $W = W(\mu/\LUV)$ and \eqref{eq:use} one gets\footnote{In general
there are also quadratic and quartic divergences which need to be subtracted by suitable counterterms. In a supersymmetric theory those divergences cancel to zero.}
\begin{equation} \label{CounterDerivative}
 \int d^4 x \sqrt{\til{\gm}} \vev{ \Rtr{{T}}{\rho}} =
 - \lim_{\LUV \to \infty}  \frac{\partial}{\partial \ln \LUV} \W_{\rm reg}= 
- \lim_{\LUV \to \infty} \Tr(e^{- \frac{  \Delta^{(0)}} {\Lambda_{UV}^2}  }) = - b_4 \;,
\end{equation}
where $b_4$ is a coefficient of the asymptotic Heat Kernel expansion 
\begin{equation} \label{HeatKernel}
 \Tr(e^{-t  \Delta^{(0)}})= \sum_{n \geq 0} b_{n}t^{\frac{n-d}{2}} \;.
\end{equation}
Using  the plane-wave basis to evaluate the trace we obtain
\begin{equation}
 \int d^4 x \sqrt{\til{\gm}} \vev{ \Rtr{{T}}{\rho}} = -\frac{1}{16 \pi^2}\frac{1}{90}  \int d^4 x \big [3 \Box^2 s -2 \Box(\partial s)^2 - 4 \partial \cdot (\partial s ((\partial s)^2
-\Box s)) \big ] \;,
\end{equation}
which decomposes into the the following invariants
\begin{equation}
\int d^4 x \sqrt{\til{\gm}} \vev{ \Rtr{{T}}{\rho}} = \frac{1}{16 \pi^2}\frac{1}{90}  \int d^4 x \sqrt{\til{\gm}}(-\frac{1}{2} \til{\Box} \til{R}+\frac{1}{4}\til{E}_4) \;,
\end{equation}
where the geometric quantities $\til{R}$ and $\til{E}_4$ are defined with respect to the 
metric $\til{g}_{\rho \la}(s)$ given above.
The result quoted in \eqref{eq:Afree0} follows by comparing the  equation above to \eqref{eq:TA}.

\section{Renormalisation group equations for $\W$}
\label{app:RGs}

In this appendix we summarise the RG equation obeyed by $\W$ and how they relate to the trace anomaly.  
For future reference and completeness  an explicit   scale symmetry breaking term in form of  
a matter mass term is added. 
The quantum vacuum transition amplitude $\W$ obeys 
an RG equation 
\begin{equation}
\label{eq:RGE}
\left(  \frac{\partial}{\partial \ln \mu} +  \beta \frac{\partial}{\partial g} -  \ga_m   \pd{\ln m }
   \right ) \W = 0 
\;, \quad \ga_m \equiv   - \pdb{\ln \mu}{\ln m} \;,
\end{equation}
which follows from $\ddb{\mu}{ \W} = 0$.\footnote{Throughout this paper $\ga = - \ga_m $ is the anomalous dimension of the squark composite operator 
 whereas other authors \cite{FO98,AKZ15}
use the anomalous dimension of the superfield $\Phi$ as $\ga = \ga_\Phi$. The relation between the two is $-2 \ga_\Phi =  \ga_m$; i.e.  $\ga|_{\text{this work}}  = 2 \ga|_{\mbox{\cite{FO98,AKZ15}} }$.
}
 Assuming a space-dependent metric,  dimensional analysis gives an equation of the form
\begin{equation}
\label{eq:dim}
\left(  \frac{\partial}{\partial \ln \mu}  +   \pd{\ln m }  +  2 \int d^4 x \ \gm^{\mu \nu}(x)\frac{\de}{\de \gm^{\mu \nu}(x)}  \right) \W = 0 \;. 
\end{equation}
Equations (\ref{eq:RGE},\ref{eq:dim}) can be combined into an RG equation with no explicit $\mu$-derivative
\begin{equation}
\left(   \beta \frac{\partial}{\partial g} - (1+ \ga_m)   \pd{\ln m } - 2 \int d^4 x \ \gm^{\mu \nu}(x)\frac{\de}{\de \gm^{\mu \nu}(x)} 
   \right ) \W = 0  \;.
\end{equation} 
The adaption of these equation to $\W_\tau$ involves replacing $\mu \to \mu e^\tau$ everywhere. 
Note, if $\tau$ is made space-dependent then $g(\mu e^\tau)$ and $m(\mu e^{\tau})$  and
 the partial derivatives are to be replaced by functional derivatives
$\frac{\partial}{\partial g} \to \int d^4 x \frac{\delta}{\delta g(x)}$  and 
$\frac{\partial}{\partial m} \to \int d^4 x \frac{\delta}{\delta m(x)}$ respectively.

A definition of the trace of the EMT is given by
\begin{equation}
\label{eq:above}
\vev{ \Rtr{{T}}{\rho}} =  - 2  \  \frac{\gm^{\mu \nu}(x)}{\sqrt{\gm(x)}}\frac{\de}{\de \gm^{\mu \nu}(x)}  \W
\end{equation}
where $\gm(x)$ denotes the determinant of the metric.
Combining \eqref{eq:above}  with \eqref{eq:dim} the following equations are obtained 
\begin{equation}
\label{eq:use}
 \int d^4 x  \sqrt{  \gm} \ \vev{ \Rtr{{T}}{\rho}}_{\rm anom} =  \frac{\partial}{\partial \ln \mu} \W   \;,  
 \quad   \int d^4 x  \sqrt{  \gm} \ \vev{ \Rtr{{T}}{\rho}}_{\rm expl} =  \frac{\partial}{\partial \ln m} \W \;,
\end{equation}
where the subscripts  ``anom" and  ``expl" refer to  anomalous and  explicit scale breaking respectively.

\bibliographystyle{utphys}
\bibliography{input}

\end{document}